# Determination of Friction Coefficient in Unconfined Compression of Brain Tissue

Badar Rashid[a], Michel Destrade[b,a], Michael Gilchrist[a*]

[a]*School of Mechanical and Materials Engineering, University College Dublin, Belfield, Dublin 4, Ireland*

[b]*School of Mathematics, Statistics and Applied Mathematics, National University of Ireland Galway, Galway, Ireland*

**Corresponding Author*

Tel: + 353 1 7161884 Fax: + 353 1 283 0534

Email: Badar.Rashid@ucdconnect.ie, michel.destrade@nuigalway.ie, michael.gilchrist@ucd.ie

**Abstract** Unconfined compression tests are more convenient to perform on cylindrical samples of brain tissue than tensile tests in order to estimate mechanical properties of the brain tissue because they allow for homogeneous deformations. The reliability of these tests depends significantly on the amount of friction generated at the specimen/platen interface. Thus, there is a crucial need to find an approximate value of the friction coefficient in order to predict a possible overestimation of stresses during unconfined compression tests. In this study, a combined experimental – computational approach was adopted to estimate the dynamic friction coefficient $\mu$ of porcine brain matter against metal platens in compressive tests. Cylindrical samples of porcine brain tissue were tested up to 30% strain at variable strain rates, both under bonded and lubricated conditions in the same controlled environment. It was established that $\mu$ was equal to 0.09 ± 0.03, 0.18 ± 0.04, 0.18 ± 0.04 and 0.20 ± 0.02 at strain rates of 1, 30, 60 and 90/s, respectively. Additional tests were also performed to analyze brain tissue under lubricated and bonded conditions, with and without initial contact of the top platen with the brain tissue, with different specimen aspect ratios and with different lubricants (Phosphate Buffer Saline (PBS), Polytetrafluoroethylene (PTFE) and Silicon). The test conditions (lubricant used, biological tissue, loading velocity) adopted in this study were similar to the studies conducted by other research groups. This study will help to understand the amount of friction generated during unconfined compression of brain tissue for strain rates of up to 90/s.

Keywords *Tissue, Computational, Ogden, Dynamic, Specimen, Lubricant, Bonded, Platen*

## 1 Introduction

Traumatic brain injury (TBI) is recognised as a leading cause of death and disability and, as such, has been the focus of extensive research for at least 50 years. The complex mechanical behavior of brain tissue due to a sudden impact on the head is still under extensive investigations. Several research groups have developed numerical models which contain detailed geometric descriptions of the anatomical features of the human head, in order to simulate and investigate internal dynamic responses to multiple loading conditions (Ho and Kleiven, 2009; Horgan and Gilchrist, 2003; Kleiven, 2007; Ruan et al., 1994; Zhang et al., 2001). However, the fidelity of these models is highly dependent on the accuracy of the material properties used to model the biological tissues.

Several research groups investigated the mechanical properties of brain tissue in order to establish constitutive relationships over a wide range of loading conditions. Mostly dynamic oscillatory shear tests were conducted over a frequency range of 0.1 to 10000 Hz (Arbogast et al., 1997; Bilston et al., 2001; Brands et al., 2004; Darvish and Crandall, 2001; Fallenstein et al., 1969; Ho and Kleiven, 2009; Hrapko et al., 2006; Nicolle et al., 2004; 2005; Ning et al., 2006; Peters et al., 1997; Prange and Margulies, 2002; Shen et al., 2006; Shuck and Advani, 1972; Takhounts et al., 1999; Thibault and Margulies, 1998) and unconfined compression tests (Cheng and Bilston, 2007; Estes and McElhaney, 1970; Franceschini et al., 2006; Miller and Chinzei, 1997; Pervin and Chen, 2009; Prange and Margulies, 2002; Tamura et al., 2007), while a limited number of tensile tests (Franceschini et al., 2006; Miller and Chinzei, 2002; Tamura et al., 2008; Velardi et al., 2006) were performed.

Unconfined compression tests are convenient to perform on cylindrical samples of brain tissue to determine mechanical properties as compared to tensile tests because they allow for homogeneous deformations. Tensile tests are typically conducted on glued cylinders, because of the fragile nature of brain tissue, which cannot easily be cut into dog bone specimens and clamped. The resulting deformation is then inhomogeneous (see Miller and Chinzei, (2002). However, the generation of undesirable friction at the specimen/platen interface is unavoidable during compression tests at quasi-static and dynamic loading conditions. Williams and Gamonpilas (2008) derived analytical solutions for the compression of cylinders with bonded surfaces and with Coulomb friction conditions at the interfaces. It was shown that the apparent moduli were strong functions of Poisson's ratio and of the Coulomb friction coefficients. A mathematical model for the realistic friction contact conditions with non-linear characteristics was also developed by Oden and Pires (1983) and Zhong (1989), which is available for use in the finite element analysis code ABAQUS 6.9/Explicit. Various research groups (Estes and McElhaney, 1970; Miller and Chinzei, 1997; Prange and Margulies, 2002; Tamura et al., 2007) have tried to reduce the effects of friction in their experimental results during unconfined compression of brain tissue. The reliability of unconfined compression tests does indeed depend significantly on the amount of friction generated during the tests. Thus, there is a crucial need to estimate an approximate value of the friction coefficient in order to quantify the overestimation of stresses during unconfined compression tests, particularly for soft biological tissue.

Therefore, in this study, unconfined compression tests were conducted on cylindrical specimens of porcine brain tissue at strain rates of 1, 30, 60, 90/s in order to estimate the friction coefficient, $\mu$ at the brain specimen/platen interface. Tests were conducted both under lubricated conditions (nearly pure slip condition) and bonded conditions (no slip condition). A combined experimental – computational approach was adopted to estimate the friction coefficient, $\mu$ at variable loading conditions. Moreover, additional tests were also performed to analyze the effects under lubricated and bonded conditions (i) with and without initial contact between the top platen and the brain specimen before the start of compression tests (ii) with different specimen aspect ratios and (iii) with different lubricants (Phosphate Buffer Saline (PBS), Polytetrafluoroethylene (PTFE) and Silicon). This study will provide further insight regarding brain tissue behavior and variation of friction coefficient, $\mu$ at different loading conditions during unconfined compression tests.

## 2 Materials and Methods

### 2.1 Specimen Preparation

Twelve fresh porcine brains were collected from a local slaughter house and tested within 3 h postmortem. Each brain was preserved in a physiological saline solution at 4 to 5$^{o}$C during transportation. Then, 24 specimens were excised from 12 porcine brains (2 specimens from each brain). The dura and arachnoid were removed and the cerebral hemispheres were first split into right and left halves by cutting through the corpus callosum. Cylindrical specimens (15.0 ± 0.1 mm diameter and 5.1 ± 0.1 mm thick) composed of mixed white and gray matter were prepared using a circular steel die cutter. The time elapsed between harvesting of the first and last specimens from each brain was 10 ~ 12 minutes for the unconfined compression tests (lubricated and bonded tests). The specimens were not all excised simultaneously, rather each specimen was tested first and then another specimen was extracted from the cerebral hemisphere. All samples were prepared and tested at a nominal room temperature of 22 $^{o}$C and relative humidity of 34 – 35%. Eighty tests on twelve fresh porcine brains were also performed the next day and were completed within 5 hours postmortem. 80 specimens were excised from 10 porcine brains (8 specimens from each brain). The specimen preparation procedure was the same as adopted for the previous tests. However, the diameter and thickness were different (diameter: 20.0 ± 0.1 mm, thickness: 6.66 ± 0.1 mm, 7.0 ± 0.1 mm 10.0 ± 0.1 mm) in order to perform tests at different aspect ratios. These tests were also useful to analyze the effectiveness of different lubricants and bonded compression.

### 2.2 Experimental Setup

A custom made test apparatus was used in order to perform unconfined compression tests at strain rates ≤ 90/s, as schematically shown in Fig. 1. An electronic actuator (speed: 500 mm/s, stroke: 50 mm, LEY 16 A, SMC Pneumatics, Ireland) was used to ensure uniform velocity during compression of brain tissue. A GSO series ± 5 N load cell (rated output of 1 mV/V nominal, Transducer Techniques, USA) was used for the measurement of compressive force. The load cell was calibrated against known masses and a multiplication factor of 13.62 N/V was used for the conversion of voltage to load. An integrated single-supply instrumentation amplifier (AD 623 G =100, Analog Devices) with a built-in single pole low-pass filter having a cut-off frequency of 10 kHz was used, with a sampling frequency of 10 kHz. The linear variable displacement transducer (LVDT) with a range ± 25 mm (ACT1000A, RDP electronics) was used to measure displacement during the unconfined compression phase.

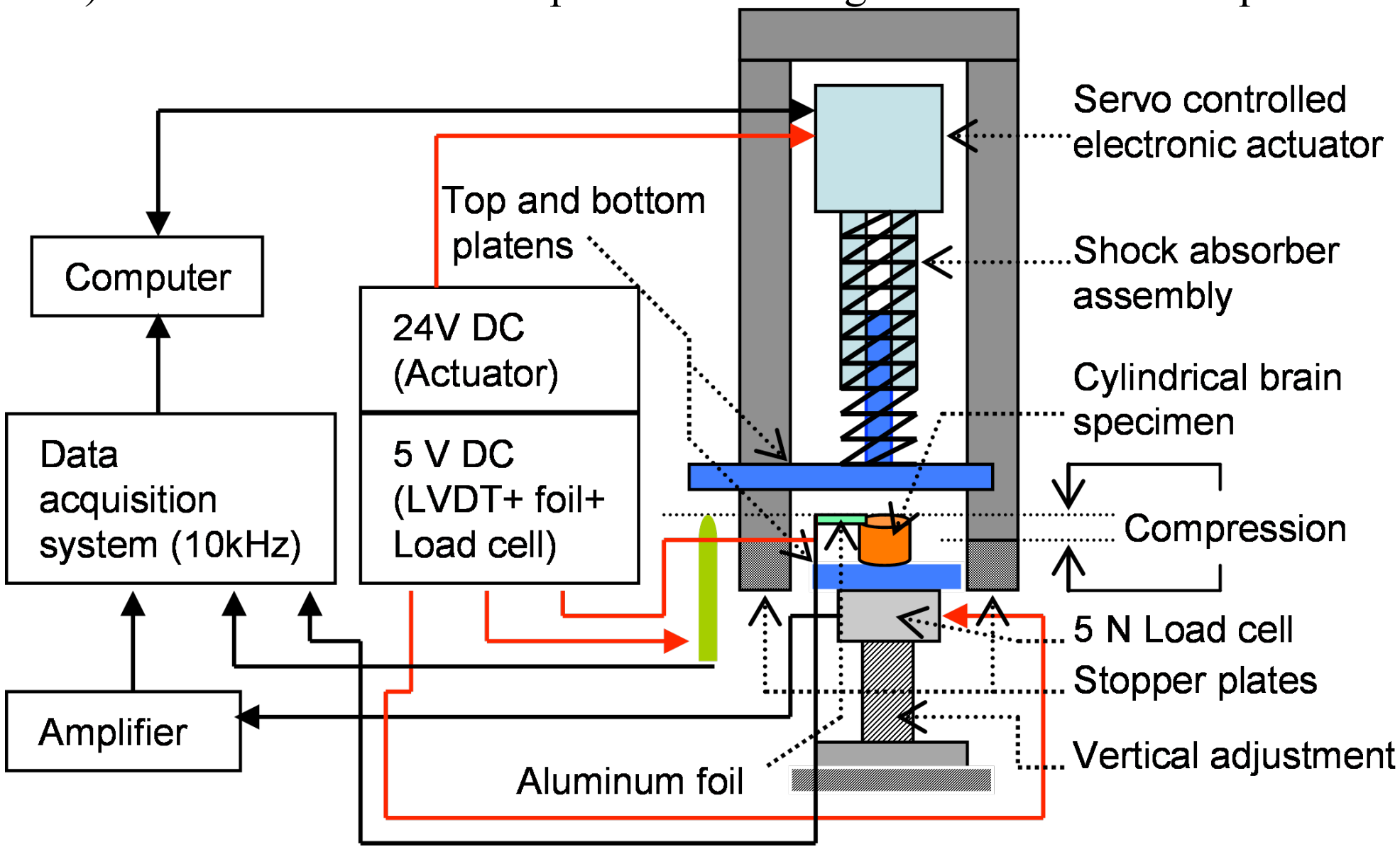

Fig. 1 – Schematic diagram of complete test apparatus for unconfined compression of brain tissue.

## 2.3 Functioning of Test Setup

A shock absorber assembly with a compression platen (top platen) and stopper plate was developed in order to avoid any damage to the electronic assembly during the experiments, as shown in Fig. 1. The stopper plate was fixed on the machine column to stop the compression platen at the middle of the stroke with an impact during the uniform velocity phase, while the actuator completes its travel without producing any backward thrust (thrust absorbed by the spring) to the servo motor and other components. Moreover, the actual actuator velocity was kept higher than the required (theoretically calculated) velocity to overcome the opposing forces (force of LVDT probe and sliding components) acting against the compression platen. During the calibration process, the actuator was run several times with and without any brain tissue specimen to ensure uniform velocity at each strain rate.

## 2.4 Lubricated and Bonded Compression

24 brain specimens were excised from 12 porcine brains (2 specimens from each brain) to perform bonded and lubricated unconfined compression tests. 6 bonded and 6 lubricated tests were performed at each strain rate of 1/s and 60/s (results at 30/s and 90/s strain rates were obtained from our previous study (Rashid et al., 2012)). Since two specimens were extracted from the same brain, one specimen was utilized for the lubricated unconfined compression tests while the other was used for the bonded unconfined compression tests. The tests were performed on mixed white and gray matter on cylindrical specimens up to 30% strain. The top and lower platens were thoroughly lubricated with Phosphate Buffer Saline (PBS) solution warmed at $37^{o}$ C to minimize frictional effects and to ensure, as much as possible, uniform expansion in the radial direction. All samples were tested at a room temperature of 22 $^{o}$C. Each specimen was tested once and then discarded because of the highly dissipative nature of brain tissue. The velocity of the compression platen (top platen) was adjusted to 300 mm/s to achieve a strain rate of 60/s using the test apparatus shown in Fig. 1. A precompression force of 0.1 mN was applied for all tests (lubricated and bonded) performed at a quasi static strain rate of 1/s; no precompression force was applied for the dynamic strain rate tests (30, 60, 90/s). A standard universal tensile testing machine (Tinius Olsen) was used in order to perform tests at a strain rate of 1/s (300 mm/min).

## 2.5 Specimen Attachment for Bonded Compression

Reliably attaching soft tissue to the platens was vital in order to achieve high repeatability. The surfaces of the platens were first covered with a masking tape substrate to which a thin layer of surgical glue (Cyanoacrylate, Low-viscosity Z105880–1EA, Sigma-Aldrich) was applied. The prepared cylindrical specimen of tissue was thoroughly lubricated with PBS solution and then placed on the lower platen. The top platen attached to the electronic actuator moved down to produce the required compression in the specimen while the lower platen sensed the force signal, which was attached to the 5 N load cell. The specimen remained hydrated during the test because of the very short time duration (4 – 8 s) of the complete test, which included placing the specimen on the lower platen and the time for the electronic actuator to compress the brain tissue.

## 2.6 Compression Test with and without Initial Contact

Two test protocols (with and without initial contact) were adopted during experimentation. All lubricated and bonded tests at a quasi static strain rate of 1/s started when the top platen was initially in contact with the specimen. Uniform velocity during the compression phase of the tissue was easily achieved at a low strain rate. As it was not possible to achieve uniform velocity at dynamic strain rates (30, 60, 90/s), all tests (lubricated and bonded) started before the platen came in contact with the brain tissue. Ten lubricated tests using PBS solution and ten bonded tests using surgical glue were performed at 0.005/s strain rate, with initial contact (WIC) and without initial contact (WOIC) in order to analyze the basic difference between the protocols.

## 2.7 Effects of Aspect Ratio

All bonded tests were performed using surgical glue. Under ideally bonded conditions and by considering that the material was incompressible, the apparent modulus was found to be dependent on the aspect ratio (diameter/height). Therefore, it was required to perform tests with different aspect ratios (15/10, 15/5) as well with the same aspect ratios (15/5, 20/6.66) to analyze the reliability of bonding during compression tests. Tests with different aspect ratios were performed at 0.005/s strain rate. Moreover, the specimens were also inspected after the test for any tear or breakage.

## 2.8 Effects of Different Lubricants

PBS solution was used in the lubricated tests to estimate friction coefficients by using combined experimental – computational approach, however it was interesting to analyze the effects of other lubricants also. Therefore, ten tests were performed by using Polytetrafluoroethylene (PTFE) lubricant (viscosity: 13 cSt at 25$^{o}$C, 1 cSt = 1 mm$^2$/s, specific gravity: 0.89 at 20$^{o}$C) and Silicon lubricant (viscosity: 10 cSt at 25$^{o}$C, specific gravity: 0.935 at 25$^{o}$C) to achieve homogeneous expansion of brain tissue at 0.005/s strain rate. These tests were essential for the comparative analysis of lubricants during unconfined compression of brain tissue (PBS, PTFE and Silicon).

# 3 Results

## 3.1 Lubricated and Bonded Tests to Estimate Friction

A combined experimental and computational approach was adopted in order to estimate the amount of friction generated during unconfined compression of brain tissue. The experimental data as discussed in Section 2.4 and 2.5 was utilized to estimate the friction coefficient. The cylindrical brain specimens were compressed up to 30% strain at different strain rates (1, 30, 60, 90/s). For a 5.0 mm thick specimen and to achieve 30% strain, the actuator velocity was adjusted to 150, 300 and 450 mm/s in order to achieve strain rates of 30, 60 and 90/s, respectively. Preliminary force - time data obtained at each strain rate was recorded at a sampling rate of 10 kHz through a data acquisition system. The force (N) was then divided by the surface area measured in the reference configuration in order to compute the compressive engineering stress (kPa). The engineering stresses determined experimentally under both lubricated and bonded conditions are shown in Fig. 2.

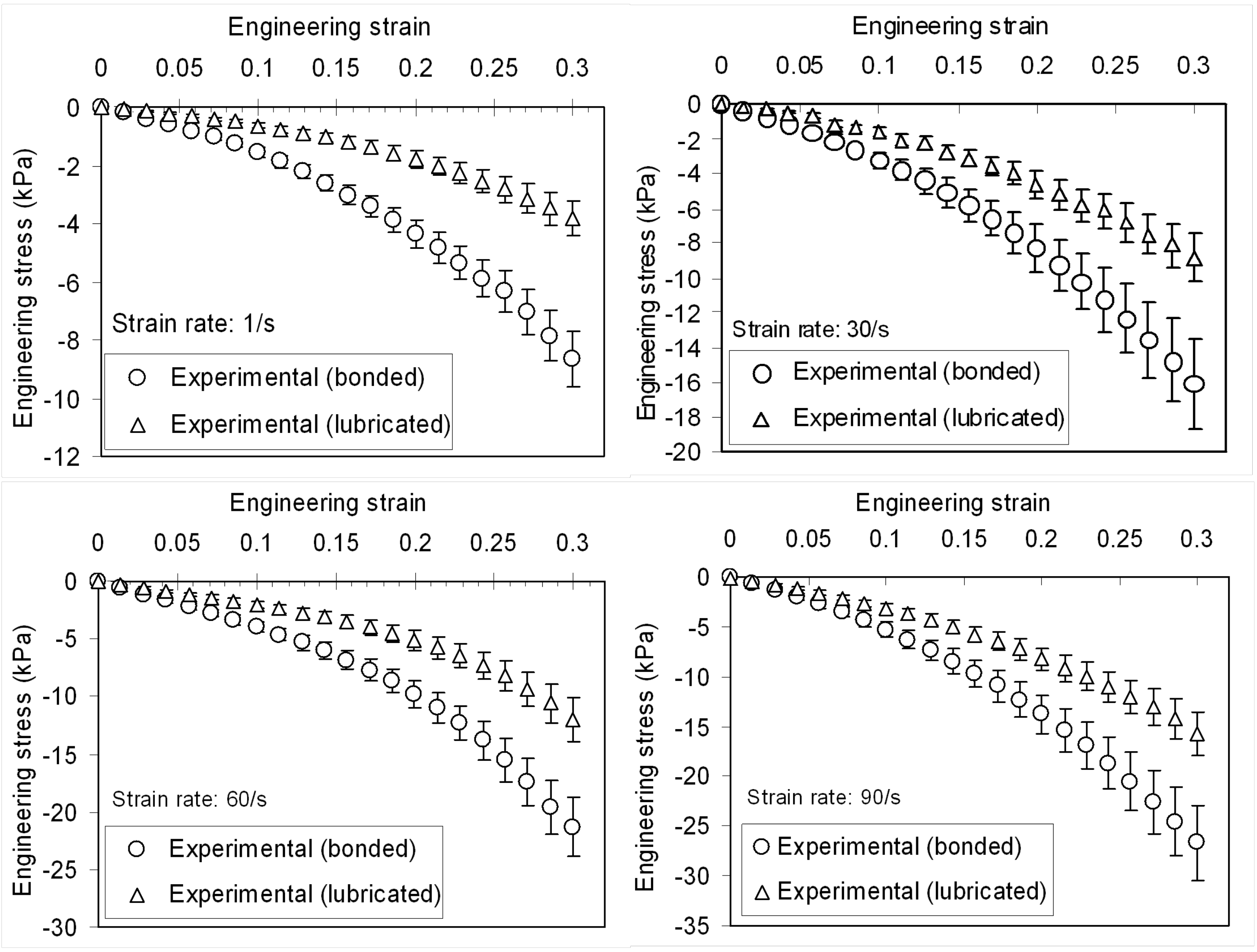

Fig. 2 – Experimental engineering stress (kPa) both in bonded and lubricated conditions at different strain rates.

### 3.2 Effects of Different Test Protocols, Aspect ratios and Lubricants

As discussed in Section 2.0, further tests were required to investigate the behavior of brain tissue when different test protocols (WIC and WOIC) were adopted. Similarly it was important to analyze the effectiveness of bonding by testing with different and the same aspect ratios and lastly the effects of different lubricants. All these tests were conduced at a quasi static strain rate of 0.005/s. Ten tests were performed at each condition and the average engineering stress (Pa) profiles were compared as shown in Fig. 3. It is observed that the engineering stress with initial contact (WIC) under lubricated test conditions is 8.4% higher than the without initial contact (WOIC) test condition at 30% strain, as shown in Fig. 3 (a). The experimental data is also analyzed statistically using one-way ANOVA test and the value of $p = 0.5840$ indicates little variation between the stress magnitudes. Similar behavior is also observed for the bonded test condition and the average engineering stress magnitude using the WIC test condition is 10% higher at 30% strain than WOIC and also observed statistically as $p = 0.2108$, as shown in Fig. 3 (b). The engineering stresses were also analyzed at different aspect ratios, AR (15/10 = 1.5 and 15/5 = 3) under bonded conditions. The magnitude of engineering stress with AR = 3 is 52.4% higher at 30% strain than the lower AR = 1.5 and the $p = 0.00112$, as shown in Fig. 3 (c). It is interesting to note that there is no significant difference between the stress magnitudes ($p = 0.8675$), with the same aspect ratios (20.0/6.66, 15/5), as shown in Fig. 3 (d). The significant difference between the engineering stress magnitudes at different aspect ratios (15/10 and 15/5) indicates that reliable bonding of brain tissue to the platen is achieved. The apparent elastic moduli $E_0$, $E_1$ and $E_2$ at strain ranges (0 – 10%), (10 – 20%) and (20 – 30%), respectively, were calculated with different aspect ratios and are given in Table 1.

Table 1 – Apparent elastic moduli, $E_0$, $E_1$ and $E_2$ of brain tissue at different aspect ratios (mean ± SD) and P = 0.00112 . Experimental data based on Fig. 3 (c)

| Aspect ratio (diameter/height) | $E_0$ (kPa) | $E_1$ (kPa) | $E_2$ (kPa) |
|---|---|---|---|
| 15/10 | 1.15 ± 0.04 | 1.65 ± 0.10 | 2.94 ± 0.18 |
| 15/5 | 2.90± 0.08 | 3.94 ± 0.20 | 6.22 ± 0.41 |

Further tests were also conducted under the same test conditions using PBS solution, PTFE lubricant and Silicon lubricant for the comparative analysis, as shown in Fig 3. (e). Little variations between the average engineering stress profiles were observed under different lubrication conditions which was also observed statistically ($p = 0.76643$) using one-way ANOVA test. It was observed that PBS solution can be used to obtain reliable results from unconfined compression tests of porcine brain tissue and may be considered equivalent to PTFE and Silicon lubricants.

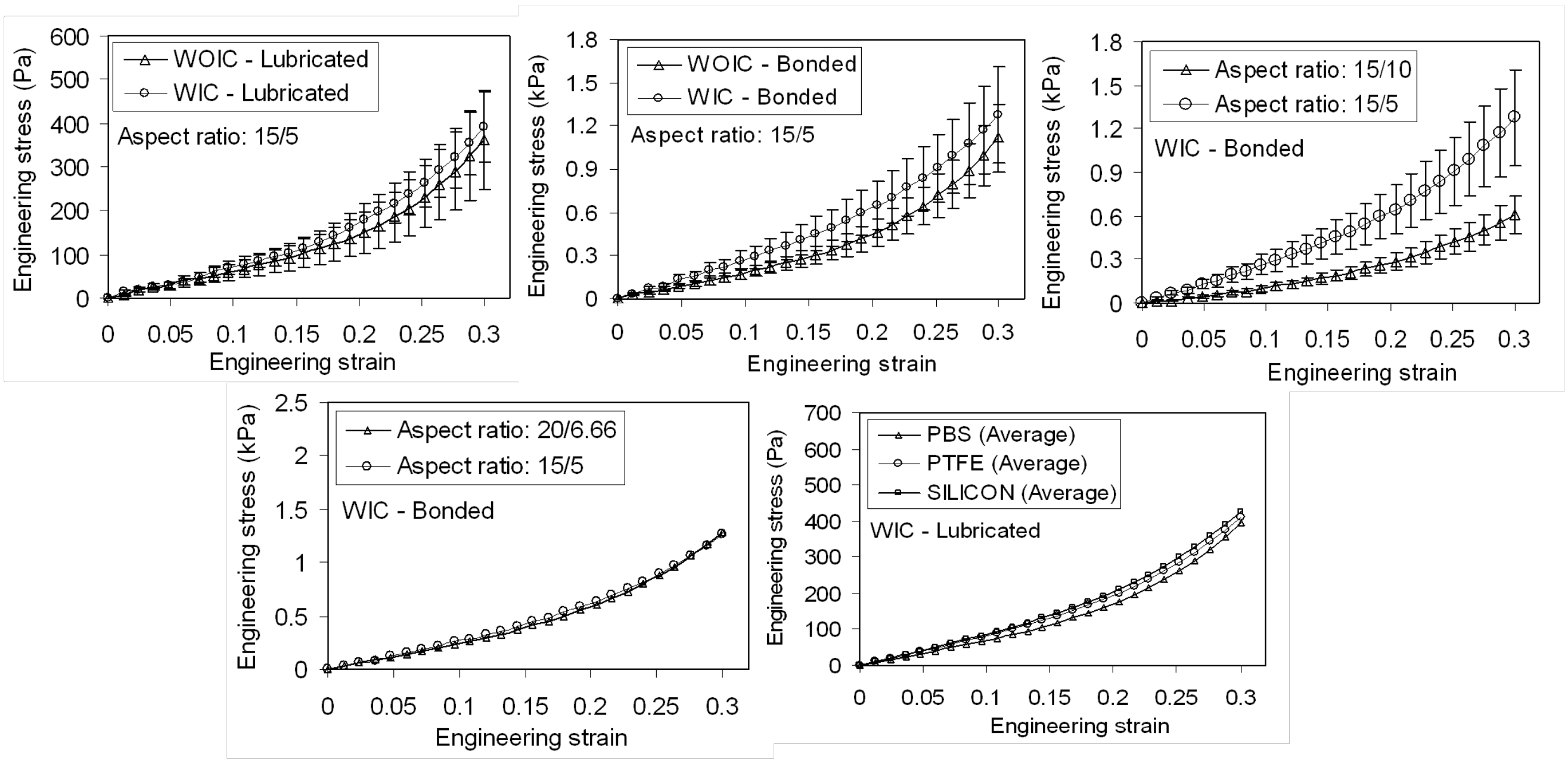

Fig. 3 – Lubricated and bonded tests performed to analyze the effects of having or not having initial contact between the top platen and the brain tissue (protocols are referred to as WIC (with initial contact) and WOIC (without initial contact)) as depicted in (a) and (b). Reliability of bonding was analyzed using different aspect ratios as depicted in (c) and (d) and the effects due to different lubricants (PBS, PTFE, Silicon) are depicted in (e). The error bars in (a), (b) and (c) correspond to ± 1 standard deviation; they have been omitted from (d) and (e) to aid with visualisation.

## 3.3 Numerical Simulations to Estimate Friction Coefficient

The experimental stress profiles were obtained both from lubricated and bonded tests, as discussed in Section 3.1. The experimental data was used as reference for the numerical simulations in ABAQUS 6.9/Explicit using Ogden material parameters (Ogden, 1972; Ogden, 1997; Ogden et al., 2004). The simulations were performed to estimate the friction coefficient. Ogden model is extensively used to characterize isotropic, nonlinear hyperelastic behavior of soft biological tissues (Brittany and Margulies, 2006; Lin et al., 2008; Miller and Chinzei, 2002; Prange and Margulies, 2002; Velardi et al., 2006). The one-term Ogden strain energy density $W$ is given by

$$W = \frac{2G}{\alpha^2}\left(\lambda_1^\alpha + \lambda_2^\alpha + \lambda_3^\alpha - 3\right) \qquad (1)$$

where $G > 0$ is the infinitesimal shear modulus (Pa), $\alpha$ is a stiffening parameter, and the $\lambda's$ are the principal stretch ratios. It yields the following nominal stress $S$

$$S = \frac{2G}{\alpha}\left\{\lambda^{\alpha-1} - \lambda^{-\left(\frac{\alpha}{2}+1\right)}\right\} \qquad (2)$$

where $\lambda$ is the stretch ratio in the direction of compression.

The experimentally determined stresses (lubricated and bonded) were used as two extreme references for the numerical simulations performed in ABAQUS 6.9/ Explicit. The density 1040 $kg/m^3$ and element *C3D8R* were used for the simulations. The *kinematic contact method*, *tangential behaviour (*as *surface interaction property)* and *penalty* option was selected to estimate the stresses at various values of friction coefficient, $\mu$. The platen surface was selected as *master* and the brain as *slave* surfaces. The numerical simulations were performed up to 30% strain. Several iterations were performed by assuming $\mu$=1 and arbitrary one-term Ogden material parameters ($G$= infinitesimal shear modulus (Pa) and $\alpha$= stiffening parameter). After a small number of simulations, excellent agreement was achieved between the bonded (experimental) and numerical stresses at $G$= 2500 Pa, $\alpha$= 2.5.

Thereafter, these material parameters were kept constant and only $\mu$ was varied until numerical stresses were in good agreement at $\mu = 0.09 \pm 0.03$ against experimental stresses at 1/s strain rate as shown in Fig. 4. A similar procedure was also adopted to estimate $\mu$ at strain rates of 30, 60, and 90/s. The numerical stress profiles at various values of $\mu$ at different strain rates are shown in Fig. 4. It is observed that $\mu$ decreases with the decrease in strain rate; however, it remains unchanged from 30 to 60/s strain rate. The estimated values of $\mu$, which are in agreement with the stress profiles obtained under lubricated conditions (Fig. 4) and the adjusted Ogden material parameters ($G$, $\alpha$) at each strain are summarized in Table 2.

Table 2. Estimated values of friction coefficients (mean ± SD)

| | | Strain rate (1/s) | | | |
|---|---|---|---|---|---|
| | | 1 | 30 | 60 | 90 |
| Friction coefficient | $\mu$ | 0.09 | 0.18 | 0.18 | 0.20 |
| | | ± 0.03 | ± 0.04 | ± 0.04 | ± 0.02 |
| Ogden parameters | $G$ (Pa) | 2500 | 5600 | 5800 | 9000 |
| | $\alpha$ | 2.5 | 2.7 | 3.3 | 3.0 |

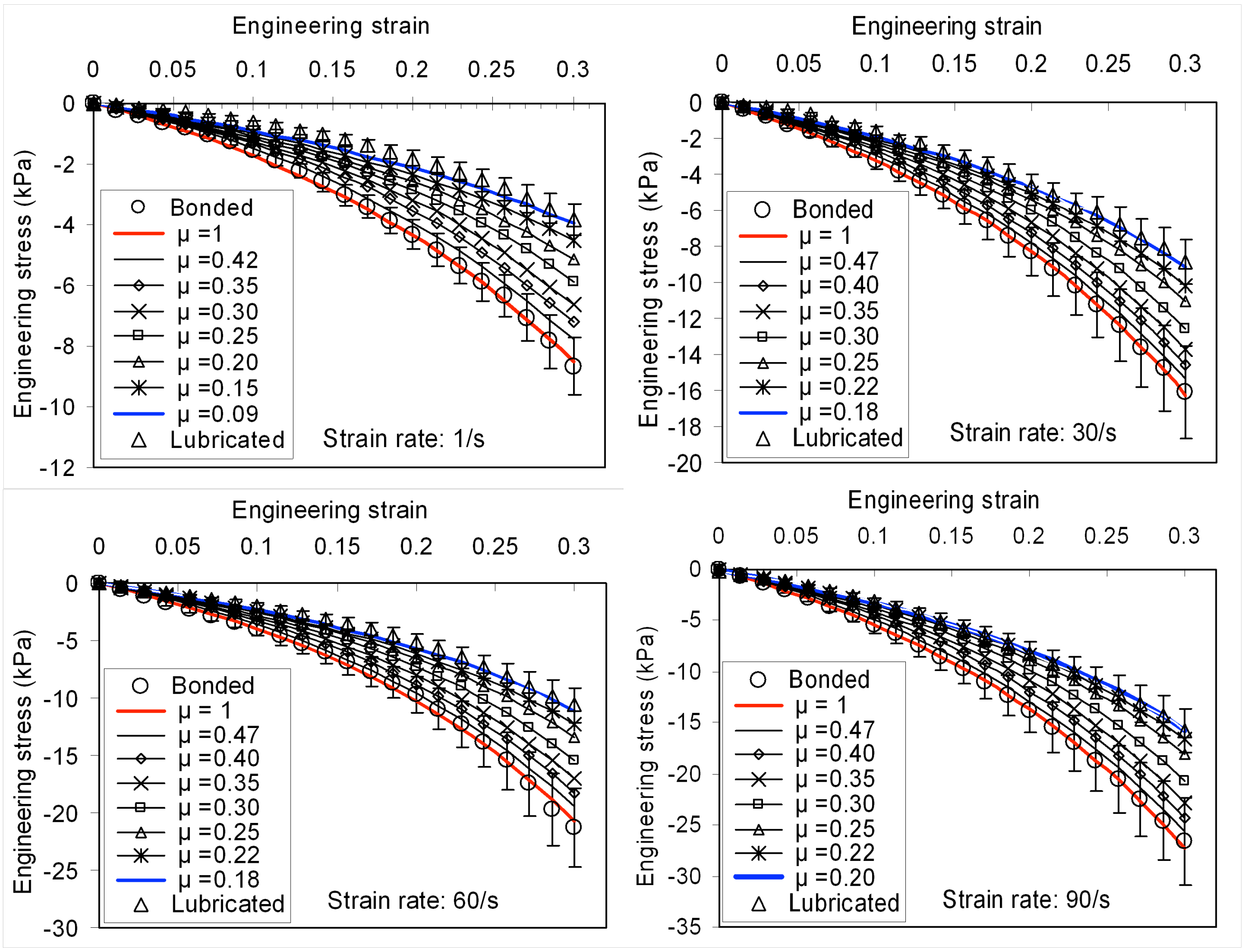

Fig. 4 – Agreement of numerical stress profiles at various values of $\mu$ with experimental stress profiles (bonded and lubricated) at various strain rates.

The $\mu$ at 30/s strain rate is 50% higher than $\mu$ at 1/s strain rate and similarly there is 11.11% increase in $\mu$ from 60 to 90/s strain rate. The range of $\mu$ varies from 0.06 (minimum) to 0.22 (maximum) over a strain rate range of 1/s to 90/s. The good agreement of numerical stresses with the experimental stresses (both lubricated and bonded conditions) was also analyzed statistically using one way ANOVA test as shown in Fig. 5

It is observed that there is no significant difference between the average experimental stress profiles (lubricated) and the numerical stress profiles (p=0.7092 ~ 0.9760) at various values of $\mu$ as shown in Fig. 5 (a). Similar agreement between the stresses is also observed under bonded condition (p=0.8721 ~ 0.9735) as shown in Fig. 5 (b).

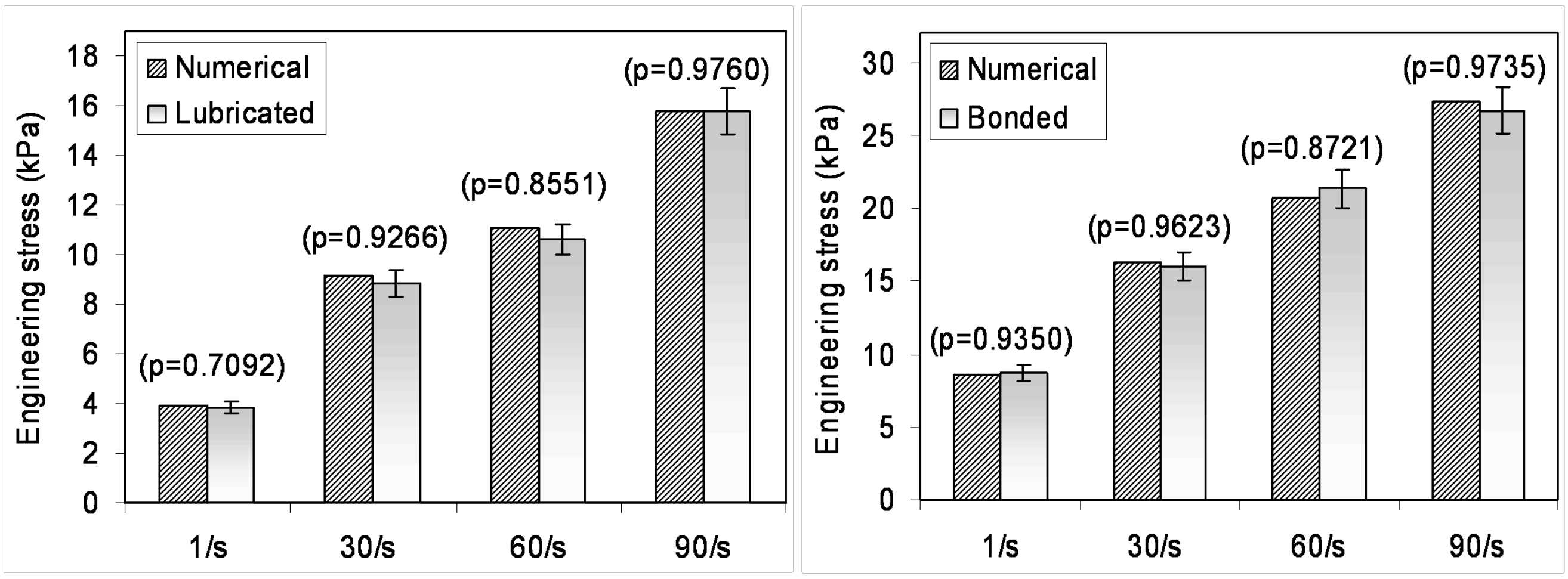

Fig. 5 – One way ANOVA tests indicates no significant difference between the numerical stresses values (obtained at various $\mu$ ) and the experimental stresses (bonded and lubricated) at various strain rates.

# 4 Discussion

Practically, it is experimentally impossible to generate a perfect frictionless condition during unconfined compression tests. Therefore, it was essential to estimate an approximate value of the friction coefficient $\mu$, particularly focusing on unconfined compression of brain tissue. A combined experimental - computational approach was adopted to determine $\mu$ at variable loading conditions. Before numerical simulations, it was essential to perform both lubricated and bonded unconfined compression tests under the same controlled environment. As discussed before, there is a significant rise in $\mu$ (50% increase) with the increase in strain rate from 1/s to 30/s. The test conditions (lubricant used, biological tissue, loading velocity) adopted in this study were similar to previous research conducted on brain tissue (Estes and McElhaney, 1970; Miller and Chinzei, 1997; Prange and Margulies, 2002; Tamura et al., 2007) as mentioned in Table 3. An approximate amount of friction which may exist during unconfined compression of brain tissue can be assumed based on the similarity of test conditions.

Table 3. Test conditions adopted by various research groups.

| Reference | Lubricant used | Biological tissue | Strain rates (1/s) |
|---|---|---|---|
| (Estes and McElhaney, 1970) | Saline at 37$^o$ C | Human and rhesus monkey | 0.0789, 0.789, 8, 40 |
| (Miller and Chinzei, 1997) | PTFE sheets | Porcine | 0.64, 0.64x10$^{-2}$, 0.64x10$^{-5}$ |
| (Prange and Margulies, 2002) | Grease (Braycote 804) | Porcine | 0.60 |
| (Tamura et al., 2007) | Saline at 37$^o$ C | Porcine | 1, 10, 50 |
| Present study | Saline at 37$^o$ C | Porcine | 1, 30, 60, 90 |

The apparent elastic modulus, $E$ (strain range: 0 – 0.10) at a strain rate of 1/s (lubricated test) is 6.35 ± 0.134 kPa (mean ± SD); this is in the same approximate range (5.7 ± 1.6 kPa at a strain rate of 1/s) as estimated by Tamura et al., (2007). Similarly, $E$ is approximately equal to 3.0 ± 0.3 kPa at a strain rate of 0.64/s, calculated from the experimental data of Miller and Chinzei (1997). However, a similar increase in $E$ of the present study was also observed at a strain range of 0.20 – 0.30 after comparing results with Tamura et al., (2007) and Miller and Chinzei (1997).

Additional experimentation was also performed to determine the reliability of bonded tests. This was done by performing tests with different aspect ratios (15/10 and 15/5). Significant variation in engineering stress magnitudes was observed (p = 0.00112) due to variation in aspect ratios, however insignificant variations were observed (p = 0.8675) with the same aspect ratios (20.0/6.66, 15/5). Therefore, it can be assumed that bonded tests performed in this study are reliable. We have also investigated the difference between the two test protocols, with initial contact (WIC) and without initial contact (WOIC), both under lubricated and bonded conditions. However it should not have any significant effect on experimental results because all tests at quasi static strain rate of 1/s were performed using the WIC test protocol and all dynamic tests were performed using the WOIC test protocol.

In the case of bonded compression up to 30% strain, the possibility of overestimation of the stress is not only due to no-slip effects but also due to 10% of the reaction force being transmitted through the expanded portion of the interface (Morriss et al., 2008). In this situation the sample planes perpendicular to the direction of the applied force do not remain plane. This aspect was thoroughly analyzed by numerical simulations following the procedure discussed by Morris et al., (2008). It was observed that 15.5% of the reaction force was transmitted through the expanded portion of the interface when the specimen planes perpendicular to the direction of the applied force do not remain plane as shown in Fig. 6.

Therefore this correction factor was included in all experimental stress profiles under bonded conditions. Moreover, Morriss et al., (2008) state that the planes perpendicular to the direction of the applied force remain plane and this assumption is invalid when the cylindrical surface of the material sample comes in contact with the compressing platen (formation of the expansion ring). In our case, 15.5% of the reaction force was transmitted through the expanded portion of the interface as the specimen planes perpendicular to the direction of the applied force do not remain plane. Therefore, the correction suggested for the vertical stretch (Miller, 2005) during unconfined compression tests is not applicable.

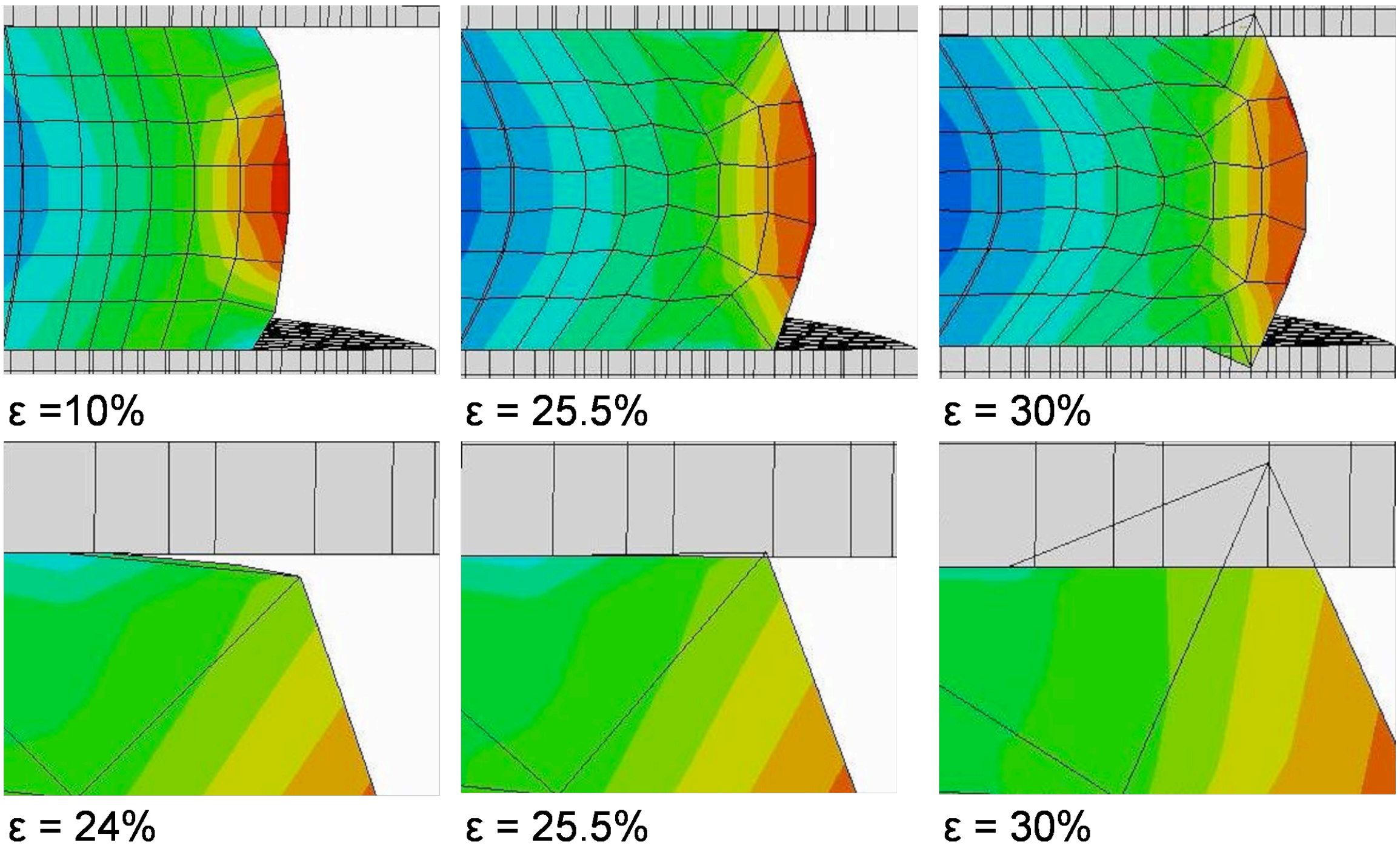

Fig.6 – Compression stages at different strain (ε) magnitudes indicating expanded portion of the interface interacting with the top platen, which transmits additional reaction force during unconfined compression of brain tissue.

The mesh density greatly reduces or increases the simulation time. The mesh is said to be convergent when further mesh refinement produces a negligible change in the solution. In the present study, mesh convergence was achieved when there was a negligible change in the numerical solution (0.3%) with further mesh refinement and the relative simulation time noted was 50 s. Moreover, the magnitudes of various energies of the numerical model were also analyzed for hourglass stiffness effects. The artificial strain energy as a percentage of the total strain energy was ~ 0.063% to 0.34% at all strain rates. The significant low percentage of artificial strain energy (≤ 0.34%) indicates that hourglassing is not a problem during simulations to determine the friction coefficient.

Wu et al., (2004), suggested that the stress levels of soft tissues can be overestimated by 10 – 50% during unconfined compression tests; however the main limitation of the study was the reference stress response curves (at zero friction) obtained after fitting to experimental data, which does include frictional effects. Thus there was a strong possibility of overestimating the real values. However, in the present study, a combined computational – experimental approach was followed to estimate values of the friction coefficient at variable strain rates. This will help to quantify any error in the magnitude of the stresses or the overestimated values of experimental stresses during unconfined compression tests.

The limitation of this study is the characterization of brain tissue as hyperelastic by using Ogden parameters; however the brain tissue is rate dependent. Nevertheless, the main purpose of this study was to estimate the amount of friction generated during unconfined compression tests, which can be achieved by following either elastic, hyperelastic or viscoelastic approaches. Therefore, a hyperelastic modelling approach was adopted for the estimation of friction coefficients, $\mu$ under different loading conditions.

In the case of dynamic tests (30, 60, 90/s strain rates) it is not possible to achieve uniform velocity if the top platen is initially in contact with the specimen. The top platen has to achieve a certain velocity before the actual compression of the tissue, as already discussed in Section 2.3. The reliability of bonding was evaluated by performing bonded tests with different aspect ratios (15/10 and 15/5). A significant dependence of the apparent elastic moduli on the different aspect ratios was observed (c.f. Table 1): this indicates that reliable bonding was achieved during the dynamic bonded tests.

During unconfined compression of brain tissue up to 30% strain, there was a possibility of shear deformation of brain tissue in the radial direction. Therefore tissue response may not be completely based on friction coefficient only but may include some part of the reaction force related to shear deformation of the tissue. However, inspection of the specimens was carried out after each bonded test and no evidence of tissue breakage / tear was observed.

The values of friction coefficients at different strains rates as mentioned in Table 2, depend only on the lubrication method used (PBS solution) and cannot be used to analyze other past results in the literature obtained by using other lubrication conditions (grease or oil). However, additional tests were also performed by using PTFE and Silicon lubricants and the experimental results were approximately similar to PBS solution. A similar approach could be adopted to determine the values of friction coefficients using grease and oil. This would provide a complete matrix of friction coefficients, which would serve to develop a friction vs strain rate law.

**Acknowledgements** This work was supported for the first author by a Postgraduate Research Scholarship awarded by the Irish Research Council for Science, Engineering and Technology (IRCSET), Ireland.

**References**

Arbogast, K.B., Thibault, K.L., Pinheiro, B.S., Winey, K.I., Margulies, S.S., 1997. A high-frequency shear device for testing soft biological tissues. J. Biomech. 30, 757–759.

Bilston, L.E., Liu, Z., Phan-Tiem, N., 2001. Large strain behavior of brain tissue in shear: some experimental data and differential constitutive model. Biorheology. 38, 335–345.

Brands, D.W.A., Peters, G.W.M., Bovendeerd, P.H.M., 2004. Design and numerical implementation of a 3-D non-linear viscoelastic constitutive model for brain tissue during impact. J. Biomech. 37, 127–134.

Brittany, C., Margulies, S.S., 2006. Material properties of porcine parietal cortex. J. Biomech. 39, 2521–2525.

Cheng, S., Bilston, L.E., 2007. Unconfined compression of white matter. J. Biomech. 40, 117–124.

Darvish, K.K., Crandall, J.R., 2001. Nonlinear viscoelastic effects in oscillatory shear deformation of brain tissue. Med. Eng. Phy. 23, 633–645.

Estes, M.S., McElhaney, J.H., 1970. Response of brain tissue of compressive loading. ASME, Paper No. 70-BHF-13.

Fallenstein, G.T., Hulce, V.D., Melvin, J.W., 1969. Dynamic mechanical properties of human brain tissue. J. Biomech. 2, 217-226.

Franceschini, G., Bigoni, D., Regitnig, P., Holzapfel, G.A., 2006. Brain tissue deforms similarly to filled elastomers and follows consolidation theory. J. Mech. Phys. Solids 54, 2592–2620.

Ho, J., Kleiven, S., 2009. Can sulci protect the brain from traumatic injury. J. Biomech. 42, 2074–2080.

Horgan, T.J., Gilchrist, M.D., 2003. The creation of three-dimensional finite element models for simulating head impact biomechanics. Int. J. Crashworthiness. 8, 353–366.

Hrapko, M., van Dommelen, J.A.W., Peters, G.W.M., Wismans, J.S.H.M., 2006. The mechanical behaviour of brain tissue: large strain response and constitutive modelling. Biorheology 43, 623–636.

Kleiven, S., 2007. Predictors for traumatic brain injuries evaluated through accident reconstructions. Stapp Car Crash J 51, 81–114.

Lin, D.C., Shreiber, D.I., Dimitriadis, E.K., Horkay, F., 2008. Spherical indentation of soft matter beyond the Hertzian regime: numerical and experimental validation of hyperelastic models. Biomech. Model. Mechanobiol. 8, 345–358.

Miller, K., 2005. Method of testing very soft biological tissues in compression. J. Biomech 38, 153-158.

Miller, K., Chinzei, K., 1997. Constitutive modelling of brain tissue: experiment and theory. J. Biomech. 30, 1115 - 1121.

Miller, K., Chinzei, K., 2002. Mechanical properties of brain tissue in tension. J. Biomech. 35, 483-490.

Morriss, L., Wittek, A., Miller, K., 2008. Compression testing of very soft biological tissues using semi-confined configuration—A word of caution. J. Biomech 41, 235-238.

Nicolle, S., Lounis, M., Willinger, R., 2004. Shear properties of brain tissue over a frequency range relevant for automotive impact situations: New experimental results. Stapp Car Crash J. 48, 239 – 258.

Nicolle, S., Lounis, M., Willinger, R., Palierne, J.F., 2005. Shear linear behavior of brain tissue over a large frequency range. Biorheology. 42, 209–223.

Ning, X., Zhu, Q., Lanir, Y., Margulies, S.S., 2006. A transversely isotropic viscoelastic constitutive equation for brainstem undergoing finite deformation. ASME J. Biomech. Eng 126, 925–933.

Oden, J.T., Pires, E.B., 1983. Nonlocal and nonlinear friction law and variational principles for contact problems in elasticity. J. Appl. Mech 50, 67–76.

Ogden, R.W., 1972. Large deformation isotropic elasticity—on the correlation of theory and experiment for incompressible rubber like solids. Proc R Soc Lond A Math Phys Sci. 326, 565–584.

Ogden, R.W., 1997. Non-linear elastic deformations. Dover, New York.

Ogden, R.W., Saccomandi, G., Sgura, I., 2004. Fitting hyperelastic models to experimental data. Comput. Mech. 34, 484 – 502.

Pervin, F., Chen, W.W., 2009. Dynamic mechanical response of bovine grey matter and white matter brain tissues under compression. J. Biomech. 42, 731–735.

Peters, G.W.M., Meulman, J.H., Sauren, A.H.J., 1997. The applicability of the time/temperature superposition principle to brain tissue. Biorheology 34, 127–138.

Prange, M.T., Margulies, S.S., 2002. Regional, directional, and age-dependent properties of the brain undergoing large deformation. J. Biomech. Eng. 124, 244–252.

Rashid, B., Destrade, M., Gilchrist, M.D., 2012. Mechanical characterization of brain tissue in compression at dynamic strain rates. J. Mech. Behavior. Biomed. Mat. In press.

Ruan, J., Khalil, T., King, A., 1994. Dynamic response of the human head to impact by three - dimensional finite element analysis. J. Biomech. Eng. 116, 44–50.

Shen, F., Tay, T.E., Li, J.Z., Nigen, S., Lee, P.V.S., Chan, H.K., 2006. Modified Bilston nonlinear viscoelastic model for finite element head injury studies. J. Biomech. Eng. 128, 797 – 801.

Shuck, L.Z., Advani, S.H., 1972. Rheological response of human brain tissue in shear. ASME, J. Basic. Eng 94, 905 - 911.

Takhounts, E.G., Crandall, J.R., Matthews, B.T., 1999. Shear properties of brain tissue using non-linear Green-Rivlin viscoelastic constitutive equation, Injury Biomechanics Research, Proceedings of the 27th International Workshop, pp. 141–156.

Tamura, A., Hayashi, S., Nagayama, K., Matsumoto, T., 2008. Mechanical characterization of brain tissue in high-rate extension. J. Biomech. Sci. Eng. 3, 263 – 274.

Tamura, A., Hayashi, S., Watanabe, I., Nagayama, K., Matsumoto, T., 2007. Mechanical characterization of brain tissue in high-rate compression. J. Biomech. Sci. Eng. 2, 115 - 126.

Thibault, K.L., Margulies, S.S., 1998. Age-dependent material properties of the porcine cerebrum: Effect on pediatric inertial head injury criteria. J. Biomech. 31, 1119-1126.

Velardi, F., Fraternali, F., Angelillo, M., 2006. Anisotropic constitutive equations and experimental tensile behavior of brain tissue. Biomech. Model. Mechanobiol. 5, 53–61.

Williams, J.G., Gamonpilas, C., 2008. Using the simple compression test to determine Young's modulus, Poisson's ratio and the Coulomb friction coefficient. Int J Solids and Structures 45, 4448–4459

Wu, J.Z., Dong, R.G., Schopper, A.W., 2004. Analysis of effects of friction on the deformation behavior of soft tissues in unconfined compression tests. J. Biomech 37, 147–155

Zhang, L., Yang, K.H., Dwarampudi, R., Omori, K., Li, T., Chang, K., Hardy, W.N., Khalil, T.B., King, A.I., 2001. Recent advances in brain injury research: a new human head model development and validation. Stapp Car Crash J. 45, 369 – 393. .

Zhong, Z.H., 1989. Contact problems with friction, Proceedings of Numiform 89, Conference, Balkema, Rotterdam pp. 599–606.